# The energetics of flat and rotating early-type galaxies and their X-ray luminosity

L. Ciotti[1] and S. Pellegrini[2,3]

[1] *Osservatorio Astronomico di Bologna, via Zamboni 33, I-40126 Bologna*
[2] *European Southern Observatory, Karl-Schwarzschild Str. 2, D-85748 Garching bei München*
[3] *Dipartimento di Astronomia, Università di Bologna, via Zamboni 33, I-40126 Bologna*

**ABSTRACT**

A recent analysis of the relationship between optical and X-ray properties for the *Einstein* sample of early-type galaxies has revealed that S0 galaxies have lower mean X-ray luminosity $L_X$ per unit optical luminosity $L_B$ than do ellipticals. In the same analysis, significant correlations are found between the X-ray properties and the axial ratios, such that the roundest systems have the highest $L_X/L_B$; this trend holds also for either Es or S0s alone (Eskridge et al. 1995a,b). The systematic X-ray underluminosity of S0s with respect to Es could be explained with a higher heat input or with a lower gravitational energy (all per unit gas mass), at fixed $L_B$. The gravitational energy could be lower because of their higher rotation, which decreases the effective potential, or their different mass distribution. These possibilities are examined here, by considering their rôle in the global energy budget of the hot gas flows in early-type galaxies. The effect of the flattening of the mass distribution is investigated with galaxy models described by the Miyamoto-Nagai potential–density pair, to which a dark matter halo of various shapes is added. For these two-component models the analytical expressions of the gravitational energy, and the stellar kinetic energy associated with various relative amounts of random motions and rotational streaming, are given. It is found that rotation cannot produce a change in the flow phase of the hot gas, independently of the galaxy shape and the presence of dark matter. The effect of flattening instead can be substantial in reducing the binding energy of the hot gas. Thus S0s and possibly non spherical Es are less able to retain a significant halo of hot gas than rounder Es of the same $L_B$.

## 1 INTRODUCTION

X-ray observations, beginning with the *Einstein* Observatory (Giacconi et al. 1979), have demonstrated that normal early-type galaxies are X-ray emitters, with 0.2–4 keV luminosities ranging from $\sim 10^{40}$ to $\sim 10^{43}$ erg s$^{-1}$ (Fabbiano 1989; Fabbiano, Kim, & Trinchieri 1992). The X-ray luminosity $L_X$ is found to correlate with the blue luminosity $L_B$ ($L_X \propto L_B^{1.8\pm0.1}$), but there is a large scatter of approximately two orders of magnitude in $L_X$ at any fixed $L_B > 3 \times 10^{10} L_\odot$. The observed X-ray spectra of the brightest objects are consistent with thermal emission from hot, optically thin gas (Canizares, Fabbiano & Trinchieri 1987). These hot gas coronae are accumulated during the galaxy lifetime from the stellar mass loss, and are heated to X-ray temperatures by the thermalization of the stellar random motions, and by the type Ia supernova explosions (hereafter SNIa). The scatter in the $L_X - L_B$ diagram has been explained in terms of environmental differences (i.e., varying degrees of ram pressure stripping due to the interaction with the intracluster or intragroup medium, White & Sarazin 1991), or in terms of different dynamical phases for the hot gas flows, ranging from winds to subsonic outflows to inflows (Ciotti et al. 1991, hereafter CDPR).

Recently Eskridge, Fabbiano & Kim 1995a,b conducted a multivariate statistical analysis of data measuring the optical and X-ray properties of the *Einstein* sample of early-type galaxies; this includes 72 ellipticals and 74 S0s, and is the largest X-ray selected sample of such galaxies presently available. Eskridge et al. 1995a show that on average S0 galaxies have lower X-ray luminosity at any fixed $L_B$ than do ellipticals (see their Fig. 8a,b,c); a comparison of the distribution functions for the two morphological subsamples indicates that the S0s have lower mean $L_X$ (at the 2.8$\sigma$ level) and $L_{\rm X}/L_{\rm B}$ (at the $3.5\sigma$ level) than do the Es. Moreovover Eskridge et al. 1995b find that galaxies with axial ratio close to unity span the full range of $L_{\rm X}$, while those with large axial ratio all have $L_{\rm X} \lesssim 10^{41}$ erg s$^{-1}$. The relationship defined by $L_{\rm X}/L_{\rm B}$ is stronger than that defined by $L_{\rm X}$, and it is in the sense that those systems with the largest $L_{\rm X}/L_{\rm B}$ are the roundest; this means that at any fixed $L_{\rm B}$ the X-ray brightest galaxies are also the roundest. This correlation holds for both morphological subsets of Es and S0s. Thus S0s and non spherical Es seem to be less able to retain hot gaseous halos than are rounder systems of the same $L_{\rm B}$. This could be due to a lower gravitational energy, because of their higher rotation, which decreases the effective potential, or their different mass distribution (see also Pellegrini 1994).

The purpose of this paper is to investigate whether a flat and partially rotationally supported galaxy is expected to host a different gas flow phase with respect to a spherical pressure supported galaxy of the same $L_{\rm B}$. This is accomplished with the study of the energetic balance of the hot gas flows. Axisymmetric two-component galaxy models are built, where the stellar and dark mass distributions have the Miyamoto-Nagai shape; the shape of the dark matter halo is varied from flat to spherical. The decoupling between ordered and disordered azimuthal motions is assumed to satisfy the Satoh's $k$-decomposition, or a "maximum rotation" decomposition. The analytical expressions for the virial terms of these two-component models, together with the projected kinematical properties for the one-component model, are given here for the first time.

In Section 2 the rôle of flattening and rotation in the energetic balance of a galaxy of a general shape is qualitatively studied. In Section 3 the two-component galaxy models are introduced; in Section 4 and 5 these are used to derive quantitative estimates of the roles of rotation and flattening. In Section 6 other possible explanations for the X-ray underluminosity of flat systems are discussed, and the main conclusions are summarized.

## 2 THE ENERGY BUDGET OF THE HOT GAS

For pressure supported spherical galaxy models, a reliable estimator of the nature of the flow phase of the hot gas can be derived from its global energy balance (CDPR). This estimator is the ratio $\chi$ between the power required to steadily extract the stellar mass loss from the galaxy potential well, and the power supplied to the gas by the thermalization of the stellar random motions and by the supernova heating. Spherically symmetric hydrodynamical simulations, in which radiation cooling is taken into account, show that in the scenario of subsequent wind, outflow, and inflow (hereafter WOI), $\chi$ describes in an accurate way the nature of the flow. The simulations show the correspondance between $\chi < 1$ and the presence of a wind phase (i.e., of a low $L_{\rm X}$), and the onset of a high $L_{\rm X}$ soon after $\chi$ becomes $> 1$ (CDPR). They also show that small variations (within 10%) of $\chi$ when it is close to unity describe large flow regime variations, from winds to inflows. The situation is obviously different in other scenarios where $\chi$ is not $< 1$ from the beginning of the flow history, i.e., if the initial flow phase is not a wind (and so a major fraction of the heating goes into radiation). The use of $\chi$ is extended here to the case of galaxies of a general shape and internal dynamics.

### 2.1 The generalized parameter $\chi$

For the needs of the following discussion, the classical scalar virial theorem for a stellar distribution of spatial density $\rho_*$ interacting with a dark matter potential $\Phi_{\rm h}$ is recalled:

$$2T + \Pi = |W|, \qquad (1)$$

where $\Pi = \int \rho_* {\rm Tr}(\sigma^2) dV$ is twice the kinetic energy associated with the stellar random motions, ${\rm Tr}(\sigma^2)$ is the trace of the velocity dispersion tensor of the stars; $T = 0.5 \int \rho_* {\rm Tr}(\overline{v}_i \overline{v}_j) dV$ is the kinetic energy associated with the stellar ordered motions; $|W| = |W_*| + |W_{\rm h}| = -\int \rho_*(0.5\Phi_* - x_k \partial\Phi_{\rm h}/\partial x_k) dV$, where $|W_{\rm h}|$ is the interaction energy of the stars with the dark halo.

The terms entering $\chi$ are now defined. The power $L_{\rm g}$ required to steadily extract from the galaxy the stellar mass loss is

$$\frac{L_{\rm g}}{\alpha(t)} = -\int \rho_*(\Phi_* + \Phi_{\rm h}) dV \equiv |U| = 2|W_*| + |U_{\rm h}|, \qquad (2)$$

where $\alpha(t) = \dot{M}_*/M_*$ is the stellar mass loss rate $\dot{M}_*$ per unit stellar mass, $M_*$ being the total stellar mass (see CDPR). The thermalization of the stellar random motions heats the gas at a power $L_\sigma$ given by

$$\frac{L_\sigma}{\alpha(t)} = \frac{1}{2} \int \rho_* {\rm Tr}(\sigma^2) dV = \frac{\Pi}{2}. \qquad (3)$$

Indicating the SNIa heating per unit time with $L_{\rm SN}$, the $\chi$ corresponding to a non rotating, totally velocity dispersion supported galaxy is:

$$\chi_\Pi = \frac{L_{\rm g}}{L_{\rm SN} + L_\sigma} = \frac{|U|}{L_{\rm SN}/\alpha + 0.5|W|}. \qquad (4)$$

Note that formally $0 < \chi_\Pi \leq 2|U|/|W|$, and that $\chi_\Pi$ increases going from a very high to a null SNIa heating.

For a fixed mass distribution (i.e., a fixed $|W|$), the amount of rotational streaming can formally vary from zero to $|W|/2$; the power $L_{\rm rot}$ related to this fraction of the total kinetic energy is given by

$$\frac{L_{\rm rot}}{\alpha(t)} = \frac{1}{2} \int \rho_* {\rm Tr}(\overline{v}_i \overline{v}_j) dV = T. \qquad (5)$$

How does $L_{\rm rot}$ enter eq. (4)? There are two extreme views. In the first one, the ordered motion is considered as a change in the effective potential experienced by the gas: in this case $L_{\rm rot}$ is to be subtracted to the numerator of (4). In the opposite view all the kinetic energy of the gas, also the ordered one, is eventually thermalized: in this case $L_{\rm rot}$ is to be added to the denominator of eq. (4). We parameterize the real behavior – which lies between the two extreme cases – by defining

$$\chi = \frac{L_{\rm g} - \gamma L_{\rm rot}}{L_{\rm SN} + L_\sigma + (1-\gamma)L_{\rm rot}} = \frac{|U| - \gamma T}{L_{\rm SN}/\alpha + 0.5|W| - \gamma T} \qquad (6)$$

with $0 \leq \gamma \leq 1$. Note that if $\gamma = 0$ then $\chi = \chi_\Pi$, i.e., if the thermalization of $L_{\rm rot}$ is complete, $\chi$ coincides with $\chi_\Pi$ even when rotation is present. When instead $\gamma = 1$, i.e., in the case of no thermalization, the effect of rotation is maximum.

In the degenerate and trivial case when rotation, dark halo, and SNIa's are absent, $\chi = 4$ from eqs. (6) and (2); the same applies when rotation is fully thermalized, and again the dark halo and the SNIa's are absent. In all cases when the supernova heating is negligible, $\chi > 1$ and an inflow is present.

## 2.2 The rôle of flattening and rotation

We consider now two galaxies of the same luminosity $L_B$, i.e., with the same SNIa heating, and the same stellar mass. One of these is the reference galaxy, a completely velocity dispersion supported system; the other galaxy is the test object, with which the effects of flattening and rotation are investigated. For this study it is useful to consider the ratio between $\chi$ of the test galaxy and $\chi_\Pi$ of the reference one, whose properties are characterized by the subscript "0":

$$\frac{\chi}{\chi_\Pi} = \left(1 - \frac{\gamma T}{|U|}\right) \times$$

$$\left[\frac{|U_0|}{|U|} + \chi_\Pi \left(\frac{|W| - |W_0|}{2|U|} - \frac{\gamma T}{|U|}\right)\right]^{-1}. \quad (7)$$

A general variation in the structure and internal dynamics of the reference galaxy can be obtained through two independent steps: the first one changes the structure from the reference case to the test one, and the second step adds the rotation. The first step implies turning $U_0$ and $W_0$ into $U$ and $W$, e.g., changing the stellar mass distribution and dark matter amount and/or distribution. From eq. (7) with $T = 0$ a dimensionless parameter is found

$$Q = \frac{|U_0|}{|U| - 0.5(|W| - |W_0|)} \quad (8)$$

whose value determines the resulting $\chi$. When, by changing the galaxy structure, $Q$ becomes $< \chi_\Pi$, then $\chi > 1$; when $Q = \chi_\Pi$, then $\chi = 1$; and when $Q > \chi_\Pi$, then $\chi < 1$.

The second step adds rotation to the modified galaxy. Using again eq. (7) with $|W| = |W_0|$ and $|U| = |U_0|$, and with $\chi_\Pi$ the resulting $\chi$ of the first step, it results that: 1) if $\chi_\Pi = 1$, also $\chi = 1$, for any $\gamma T$; 2) if $\chi_\Pi > 1$, also $\chi > 1$, and an increase in $\gamma T$ produces an increase in $\chi$; 3) if $\chi_\Pi < 1$ then $\chi < 1$, and an increase in $\gamma T$ produces a decrease in $\chi$. In other words, the effect of rotation is never that of changing a $\chi > 1$ into a $\chi < 1$, and viceversa, independently of the unknown value of $\gamma$. A qualitative explanation of this behavior can be found by inspecting eq. (6): when the gas in absence of rotation is bound, the introduction of rotation causes the subtraction of the same quantity from the numerator and the denominator, and this corresponds to a larger percentual decrease of the heating than of the binding energy. So, an increase in $\chi$ is produced. The reverse is true when the gas is unbound in absence of rotation. As a consequence, in the WOI scenario, rotation cannot produce a change in the flow phase; instead, it acts in the sense of making even more stable the existing flow phase, for any fixed galaxy structure. It would be interesting, for the purpose of the present study, to know under which conditions and how frequently, at $L_B$ fixed, the flattening produces $Q > \chi_\Pi$, where $\chi_\Pi$ refers to a spherical galaxy and is $> 1$; in this case in fact $\chi$ would be $< 1$ for the flattened object, and this could host an outflow. If also some rotation is added, a further decrease in $\chi$ is produced (if $\gamma > 0$), and the flow is even more likely to be an outflow or a wind. In this way the average X-ray underluminosity of flat systems at fixed $L_B$ would be explained. What is the efficiency of the flattening in producing a variation in $\chi$, for realistic galaxy models? What is the importance of the further variation that can be produced by rotation? These points will be investigated in the case of S0s in Sections 4 and 5, using the galaxy models described below.

**Figure 1.** The variation of $\chi$ produced by structural changes. Top panel: no dark matter. In the middle and lower panels, for the same $\chi_\Pi$ values, solid lines correspond to $s = 0$, dotted lines to $s = 1$, and dashed lines to $s = 5$. In the middle panel the dark matter keeps its spherical shape while the stellar mass is flattened; in the lower panel $s_h = 0.5s$.

## 3 THE MODELS

The stellar density distribution of an S0 galaxy is modeled by adopting the potential–density pair introduced by Miyamoto & Nagai (1975), as given in Binney & Tremaine (1987, p.44; hereafter BT87):

$$\rho_* = \left(\frac{M_* b^2}{4\pi}\right) \frac{aR^2 + (a + 3\zeta)(a + \zeta)^2}{\zeta^3 [R^2 + (a + \zeta)^2]^{5/2}}, \quad (9)$$

$$\Phi_* = -\frac{GM_*}{\sqrt{R^2 + (a + \zeta)^2}}, \quad (10)$$

where $\zeta = \sqrt{z^2 + b^2}$, and $(R, \varphi, z)$ are the cylindrical coordinates. For $a = 0$ the stellar density is a Plummer (1911) sphere with effective radius $b$; for $b = 0$ eq. (9) describes the Kuzmin (1956) disk. A constant stellar mass-to-light ratio is assumed throughout the system. The Miyamoto-Nagai (hereafter MN) model produces a realistic gross structure, similar to that observed for S0s; particularly, the light distribution resembles that of disk galaxies when $a/b = 5$ (e.g., BT87). In order to study the effect of a dark matter halo, another Miyamoto-Nagai density distribution characterized by $M_h$, $a_h$, and $b_h$, is added to the stellar profile.

The quantities entering $\chi$ are determined using the Jeans equations. Following Binney, Davies, & Illingworth (1990) we implicitly adopt a distribution function that depends only on the isolating integrals $E$ and $L_z$. In this case: 1) the terms $\overline{v_R v_z}$, $\overline{v_R v_\varphi}$, and $\overline{v_\varphi v_z}$ are zero (and so the velocity dispersion tensor is aligned with the coordinate system);

2) the radial and axial velocity dispersions are equal, i.e., $\sigma_R = \sigma_z \equiv \sigma$; 3) the only nonzero streaming motion is in the azimuthal direction ($\varphi$). The resulting Jeans equations are:

$$\frac{\partial \rho_* \sigma^2}{\partial z} = -\rho_* \frac{\partial \Phi_{\rm tot}}{\partial z}, \qquad (11)$$

and

$$\frac{\partial \rho_* \sigma^2}{\partial R} + \rho_* \left( \frac{\sigma^2 - \overline{v_\varphi^2}}{R} + \frac{\partial \Phi_{\rm tot}}{\partial R} \right) = 0, \qquad (12)$$

where $\Phi_{\rm tot} = \Phi_* + \Phi_{\rm h}$.

Since in eq. (12) only $\overline{v_\varphi^2}$ appears, the decomposition of the azimuthal velocity can be chosen freely. Being the stellar density flat, a natural choice is the $k$-decomposition introduced by Satoh (1980):

$$\overline{v_\varphi}^2 = k^2 (\overline{v_\varphi^2} - \sigma^2), \qquad (13)$$

and then

$$\sigma_\varphi^2 \equiv \overline{v_\varphi^2} - \overline{v_\varphi}^2 = k^2 \sigma^2 + (1 - k^2) \overline{v_\varphi^2}, \qquad (14)$$

with $0 \leq k \leq 1$. In the case $k = 0$ all the flattening is due to the azimuthal velocity dispersion, i.e., no net rotation is present; $0 < k \leq 1$ is a sufficient condition in order to have a positive value of $\sigma_\varphi^2$ everywhere, and $k = 1$ corresponds to the isotropic rotator. This solution does not describe the maximum rotation formally allowed for a given galaxy structure; this is obtained by relaxing the condition that $k$ is a constant, and requiring that $\sigma_\varphi^2 = 0$ over the whole space:

$$k_{\max}^2(R, z) = 1 + \frac{\sigma^2}{\overline{v_\varphi^2} - \sigma^2}. \qquad (15)$$

While certainly such a solution is not physically possible for a realistic structure, nevertheless it is used as a limiting case in the investigation of the maximum effect of rotation.

## 4 MODELS WITH NO DARK HALO

In the case of no dark halo, the Jeans equations can be solved analytically (see Nagai & Miyamoto, 1976). The velocity dispersion $\sigma^2$ is recovered from (11):

$$\rho_* \sigma^2 = \frac{G M_*^2 b^2}{8\pi} \frac{(a + \zeta)^2}{\zeta^2 [R^2 + (a + \zeta)^2]^3}, \qquad (16)$$

and after some easy algebra one obtains from (12):

$$\rho_* (\overline{v_\varphi^2} - \sigma^2) = \frac{G M_*^2 a b^2}{4\pi} \frac{R^2}{\zeta^3 [R^2 + (a + \zeta)^2]^3}. \qquad (17)$$

Using eqs. (15), (16), and (17) we can recover

$$k_{\max}^2(R, z) = 1 + \frac{\zeta (a + \zeta)^2}{2 a R^2}. \qquad (18)$$

The analytical expressions for the various terms entering the virial theorem have been derived, both for the constant $k$ case and the maximum rotation case (see Appendix B).

### 4.1 The effect of flattening and rotation

We compute here the $\chi$ variation produced by the flattening and the rotation for MN models; the galaxy total luminosity is kept constant. As outlined in Section 2.2, the effect of these two processes can be considered independently, and we start examining the effect of flattening onto a spherical reference galaxy. Two structural parameters describe the process of flattening: $s = a/b$, and $r = b/b_0$, where $b_0$ is the effective radius of the reference galaxy. $s$ measures the intrinsic shape of the flattened galaxy, and $r$ measures its size.

The parameters $s$ and $r$ enter eq. (7) only through $Q$, and the effect of flattening is given by

$$\frac{\chi}{\chi_\Pi} = \frac{4 - Q}{3Q + \chi_\Pi (1 - Q)}. \qquad (19)$$

Note that, in case of no dark matter, $Q = 4/(1 + 3 L_{\rm g}/L_{\rm g,0})$: so $Q$ is directly related to the variation in the potential energy of the hot gas. Figure 1 (upper panel) shows $\chi$ vs. $Q$, for three chosen values of $\chi_\Pi$. A change in the galaxy structure is able to produce large variations in $\chi$, of the order of those needed for a change in the flow phase. More precisely, since for spherical galaxy models $\chi_\Pi$ hardly exceeds 1.5 (CDPR), at most $Q$ of this order is required to have $\chi < 1$.

The relationship between $Q$ and $(s, r)$ is shown in Fig. 2, where the contour levels of constant values for $Q(s, r)$ are plotted. The interesting region in the $(s, r)$ plane is that in which $Q(s, r) > 1$: only in this case a system with $\chi_\Pi > 1$ might become characterized by $\chi < 1$ after a flattening, if this produces $Q > \chi_\Pi$. In other words, given a spherical galaxy with $\chi_\Pi > 1$, all $(s, r)$ pairs lying in the upper part of the plane which has as lower bound the curve $Q(s, r) = \chi_\Pi$ describe flattened galaxies with $\chi < 1$. It is interesting to note that an increase in $Q$ (i.e., a decrease in $L_{\rm g}$) for roughly spherical galaxies is more easily obtained through their flattening than through an increase in their size, while the reverse is true for disky objects.

In order to find the location of real galaxies in the $(s, r)$ plane, in Fig. 2 two families of lines are shown: the first one describes the ratio of the projected central velocity dispersion $\sigma_{\rm p}$ of the flattened system seen face-on to the same quantity for the corresponding spherical reference galaxy [eq. (A15)]; the second family describes the same ratio using the edge-on velocity dispersions [the choice of the decomposition for the azimuthal component of the velocity dispersion cannot affect the central value of the projected velocity dispersion, as can be checked with eqs. (A8), (A9) and (A12)]. From Fig. 2 two important results are derived: 1) reasonable values for $r$ and the $\sigma_{\rm p}$ ratio (i.e., near unity), and for $s$ ($\lesssim 1$), correspond to $Q$ values that can produce $\chi$ variations of the order of those required to have a change in the flow phase; 2) the region where $Q > 1$ is also the region where the flattened systems have a lower central velocity dispersion than the spherical systems of the same luminosity.

A qualitative explanation for the behavior of $Q$ and the central velocity dispersion can be given. The projected velocity dispersion is proportional to the average of the potential over the stellar density, and, as a consequence, at the zeroth-order the central $\sigma_{\rm p}$ contains information on the value of the central potential. It follows that $\sigma_{\rm p} \propto \sqrt{-\Phi_*(0)} \propto 1/\sqrt{r(1+s)}$. Similarly, $L_{\rm g} \propto -\rho_*(0) \Phi_*(0) \propto (3+s)/r^4(1+s)^4$, and so increasing $s$ and $r$ lowers $\sigma_{\rm p}$ and much more $L_{\rm g}$. This qualitatively explains why $Q$ variations that are very effective in producing a change in the flow phase correspond to small changes in

**Figure 2.** Contour levels of constant $Q$ values (solid lines), in the case of no dark matter. Dashed lines show the loci where the ratio between the edge-on $\sigma_p$ of the flattened and the spherical system is 1, 0.75, 0.5; dotted lines describe the same ratios for the face-on projection.

$\sigma_p$. Eskridge et al. 1995b find a significant trend of the axial ratios with $\sigma_p$, in the sense that galaxies with the largest $\sigma_p$ are the roundest, for constant $L_B$. So, a scenario in which $\chi$ is preferentially $< 1$ in flat systems can explain their systematic X-ray underluminosity, and finds support in the optical data.

As outlined at the end of Section 2.1, the previous results apply also when rotation is fully thermalized. The efficiency of rotation in producing a variation in $\chi$ is investigated now for the limiting case of no thermalization ($\gamma = 1$), i.e., when its effect is maximum. Two cases are considered: the isotropic rotator, $T = T_{\rm is}$ [$k = 1$ in (B9)], and the maximum rotation solution, $T = T_{\rm max}$, given by (B11). Figure 3 (upper panel) shows the variation in $\chi$ produced by rotation, for a fixed galaxy structure characterized by various degrees of flattening. This variation is always very small: for example if, after a flattening with $s = 5$, $\chi_\Pi = 0.9$, then the additional reduction in the case of the maximum rotation solution is 3%. So, not only rotation cannot change the gas flow regime, but its effect is almost negligible anyway.

## 5 MODELS WITH DARK HALO

The analysis of the previous section is extended here to the two-component MN models. For a given stellar structure, the parameters defining the dark halo are: $\mathcal{R} = M_h/M_*$, $b_h$, and $s_h = a_h/b$, and so the most general change in the total mass distribution involves five parameters. Note that the freedom in the choice of $a_h$ permits to change the flattening of the dark mass independently of that of the stellar mass. In order to reduce the number of free parameters, it is assumed that $b_h = b$, which allows to perform the calculations analytically; moreover, $\mathcal{R}$ is kept constant during the process of flattening, i.e., round and disky objects of the same $L_B$ have the same amount of dark matter. Finally, it is assumed that the dark matter dominates the dynamics, i.e., in eqs. (7) and (8) only the contribution of the dark matter potential is considered. So, three parameter values are to be specified: $r, s, s_h$.

In Appendix B the analytical formulae for the terms entering the virial theorem and $\chi$ are given, for the Satoh's and the maximum rotation decomposition of the azimuthal velocity.

**Figure 3.** The variation of $\chi$ produced by rotation; $\chi$ and $\chi_\Pi$ refer to a galaxy with the same structure. The various lines correspond to various degrees of the flattening for the stellar mass ($s = 0, 1, 5$), and larger variations correspond to larger $s$. In the upper panel there is no dark matter, in the middle the dark halo is spherical, in the bottom $s_h = 0.5s$.

### 5.1 The effect of flattening and rotation

As previously done in the case of no dark matter, we start studying the trend of $\chi$ with the flattening:

$$\frac{\chi}{\chi_\Pi} = \frac{4 - Q}{(4 - \chi_\Pi)Q + 2(\chi_\Pi - Q)\widetilde{W}_h(c,s)/\widetilde{U}_h(c)} \quad (20)$$

where $c = (s + s_h)/2$, and the functions $\widetilde{W}_h$ and $\widetilde{U}_h$ are given in (B3) and (B6). In Fig. 1 we plot $\chi$ versus $Q$ for three fixed values of $\chi_\Pi$ and $s$. Concerning the shape of the dark halo, two cases are investigated: in the first one the dark matter distribution is spherical for any flattening of the visible matter ($s_h = 0$), in the other the process of flattening affects also the halo mass distribution ($s_h = 0.5s$). The case $s_h = s$ coincides with the case of a one-component MN model. Figure 1 shows that the presence of the dark matter does not change significantly the trend shown previously: the same values of $Q$ produce essentially the same variations in $\chi$ that were produced in the case of no dark matter. The relationship between $Q$ and $(s, r)$ is very similar to that obtained in the no dark matter case (see Fig. 4).

Again we overplot in the $(s, r)$ plane the contours of constant value for the ratio of the face-on $\sigma_p$'s in the flat and in the spherical reference galaxy. When $s_h = 0.5s$, conclusions very similar to those derived without dark matter are valid. As before, the $Q$ contours have roughly the same shape of the $\sigma_p$ contours, but the relative position of the two has changed slightly: now the $\sigma_p$ contours are shifted upwards with respect to the $Q$ contours. As a result, for example, the line of equal value for $\sigma_p$ in the spherical and in the flat system lies above the $Q = 1$ line, and so even flat systems with $\sigma_p$ comparable to that of the spherical galaxy can have $\chi < 1$.

**Figure 4.** Contour levels of constant $Q$ values, in the case of dominant dark matter (solid lines). In each panel the dotted lines show the loci where the ratio between the face-on $\sigma_{\rm p}$ of the flattened and the spherical system is 1.25, 1, 0.75.

$\sigma_{\rm p}$ shows instead a clearly different behavior from the case without dark matter, when the dark halo is kept spherical (Fig. 4, left panel). Note in particular that high $Q$ values can be reached while keeping $\sigma_{\rm p}$ similar in the spherical and in the test galaxy. A qualitative explanation of the behavior of $Q$ and $\sigma_{\rm p}$ follows from considerations similar to those of Section 4.1. When the dark matter dominates, $\sigma_{\rm p} \propto \sqrt{-\Phi_{\rm h}(0)} \propto 1/\sqrt{r(1+s_{\rm h})}$, and the potential energy of the gas $L_{\rm g} \propto -\rho_*(0)\Phi_{\rm h}(0) \propto (3+s)/r^4(1+s)^3(1+s_{\rm h})$. If $s_{\rm h} = 0.5s$, increasing $s$ lowers $\sigma_{\rm p}$ and $L_{\rm g}$ in the same way as in the no dark matter case. On the contrary if $s$ increases when $s_{\rm h}$ and $r$ are fixed, the $\sigma_{\rm p}$ ratio remains approximately constant, but the potential energy of the gas decreases with $\rho_*$, and so the escape of the gas is favored.

The effect of rotation is shown in Fig. 2, again for the isotropic rotator and for the maximum rotation solution (with $\gamma = 1$), for various flattening of the luminous matter. The variation in $\chi$ is slightly larger than in the case of no dark matter, especially when the dark halo is spherical. In any case, this variation for $\chi_{\rm II} \simeq 1$ is always very small, and so the result of the case of no dark matter is confirmed: the effect of rotation is negligible in the energy balance of hot gas flows.

## 6 DISCUSSION AND CONCLUSIONS

In this paper we investigate which flow phase of the hot gas is most likely in stellar systems characterized by various degrees of flattening and rotation, for the WOI scenario. Energetic arguments have been used that involve the ratio $\chi$ between the energy required to extract the hot gas from the galaxy potential well and the energy supplied for this process. For a general stellar system, we proved analytically that the critical variations in $\chi$ (i.e., from $> 1$ to $< 1$ or viceversa) can be produced only by a change in the galaxy structure; rotation – if not fully thermalized – acts in the sense of increasing a $\chi > 1$ and decreasing a $\chi < 1$.

Realistic galaxy models for S0s – with and without dark matter – have been built by using the Miyamoto-Nagai density distribution, and by comparing their central stellar velocity dispersion $\sigma_{\rm p}$ with that of a spherical galaxy of the same $L_{\rm B}$. The analytical formulae for the virial properties of the two-component galaxy models have been derived, for the Satoh's $k$-decomposition and for the maximum rotation solution, together with the projected edge-on ordered and random stellar velocity field, in the case of no dark matter.

It is shown that reasonable flattenings of the galaxy structure at fixed $L_{\rm B}$ can change a $\chi > 1$ of a round system into a $\chi < 1$, i.e., can make the gas less bound (for a $\sigma_{\rm p}$ of the flat system comparable to that of the round one). The rôle of rotation for the adopted galaxy models has proved to be negligible, because it changes $\chi$ by only a few percent even in the maximum rotation case. So, the problem of the unknown amount of thermalization of the ordered motions does not affect our conclusions.

The efficiency of the flattening and the inefficiency of rotation in changing the nature of the gas flow phase is consistent whith the observational result that – at fixed $L_{\rm B}$ – $L_{\rm X}/L_{\rm B}$ is on average lower not only in S0 galaxies with respect to Es, but also in *flat ellipticals* with respect to round ones; evidence of this trend has in fact been found by Eskridge et al. (1995b).

Other possibilities could be imagined to explain the X-ray underluminosity of S0s with respect to Es. One could be a higher SNIa rate, due to the possible presence of a younger stellar population in the disk: this could favor the escape of the gas. This solution does not seem likely, in view of the fact that also flat Es show the same trend, and it cannot be sustained that the stellar population of flatter Es has different properties from that of rounder Es. Similarly, the possible presence of an X-ray absorbing cold material in the disk of S0s does not seem to be a general explanation of the effect.

We conclude pointing out a problem with the purely energetic approach of the present investigation: $\chi$ is a global parameter, and so it does not give much indication when there is a sort of "decoupling" in the flow between different regions of the galaxy. This is the case, for example, of the partial wind phase (see, e.g., Pellegrini & Fabbiano 1994). In connection with this point we also remark that the reliability of $\chi$ has been proved by CDPR just in the case of spherical models, and the above mentioned "decoupling" could be more likely for flattened systems. Hydrodynamical simulations are required for a definite answer, and an extension of the CDPR study to axisymmetric models is in progress.

## ACKNOWLEDGEMENTS

We would like to thank P. Eskridge and G. Fabbiano for showing us their results in advance of publication. Sixty per cent of the work of L.C. was supported by the Italian Ministry of Research (MURST).

## APPENDIX A: PROJECTED PROPERTIES OF MN MODELS

In this Appendix the analytical expressions for the projected velocity dispersion associated with a stellar model described by the density-potential pair (9)-(10) are given. The projection of the velocity dispersion is performed both for the Satoh and the maximum rotation decomposition.

We start evaluating the projected surface density. Defining

$$F(s) = \begin{cases} \text{arcos}(s)/\sqrt{1-s^2} & \text{if } 0 \leq s < 1; \\ 1 & \text{if } s = 1; \\ \text{arcch}(s)/\sqrt{s^2-1} & \text{if } s > 1 \end{cases} \quad (A1)$$

the face-on projection in the center is given by $\Sigma_*(0) = (M_*/2\pi b^2)\widetilde{\Sigma}_*(s)$, where

$$\widetilde{\Sigma}_*(s) = \frac{2 + s^2 - 3sF(s)}{(1-s^2)^2}, \quad (A2)$$

with $\widetilde{\Sigma}_*(1) = 2/5$. $\Sigma_*(R)$ could not be determined analytically outside the center.

The edge-on projection of the density at the point $(R, z)$ in the projection plane is given by

$$\Sigma_*(R,z) = \frac{M_* b^2}{2\pi} \frac{aR^2 + (a+2\zeta)(a+\zeta)^2}{\zeta^3 [R^2 + (a+\zeta)^2]^2}, \quad (A3)$$

where $\zeta = \sqrt{z^2 + b^2}$. This expression was already obtained by Satoh (1980), and is given here for completeness.

Moving to the projection of the squared velocity, this is written as

$$\Sigma_* v_p^2 = \int_{-\infty}^{\infty} \rho_* \overline{\langle \mathbf{v}, \mathbf{n} \rangle^2} \, dl, \quad (A4)$$

where $\langle, \rangle$ is the inner product, $\mathbf{n}$ is the l.o.s. direction, $l$ is a parameterization of the l.o.s. integration path, and the overline represents the mean over the phase space (see, e.g., BT87, p.208). Expanding (A4) one obtains $v_p^2 = V_p^2 + \sigma_p^2$, where $V_p^2$ is the projection of the square of the component along $\mathbf{n}$ of the ordered velocity, and $\sigma_p^2$ is the projected velocity dispersion. Note that $\sigma_p^2$ is *not* the line-of-sight (i.e., the observed) velocity dispersion if a net rotation is present; this is instead given by:

$$\Sigma_* \sigma_{\text{los}}^2 = \int_{-\infty}^{\infty} \rho_* \overline{(\langle \mathbf{v}, \mathbf{n} \rangle - v_{\text{los}})^2} \, dl = \Sigma_*(v_p^2 - v_{\text{los}}^2), \quad (A5)$$

where

$$\Sigma_* v_{\text{los}} = \int_{-\infty}^{\infty} \rho_* \overline{\langle \mathbf{v}, \mathbf{n} \rangle} \, dl. \quad (A6)$$

The analytic solution of the integration in (A6) could not be found. Note however that for the *central* face-on and edge-on projections it results $\sigma_{\text{los}}^2 = \sigma_p^2$.

We now evaluate the two components of $v_p^2 = V_p^2 + \sigma_p^2$ for an edge-on projection. The integration in (A4) is made by changing the angular coordinates of $\mathbf{n}$ to the radial coordinate; in the case of the Satoh decomposition, using (13) one obtains

$$V_p^2(R,z) = k^2 \mathcal{V}_{\text{is}}^2(R,z), \quad (A7)$$

and

$$\sigma_p^2(R,z) = \mathcal{S}_{\text{is}}^2(R,z) + (1-k^2)\mathcal{V}_{\text{is}}^2(R,z), \quad (A8)$$

where the meaning of the two new functions $\mathcal{S}_{\text{is}}^2$ and $\mathcal{V}_{\text{is}}^2$ is evident. After integration

$$\Sigma_* \mathcal{V}_{\text{is}}^2 = \frac{GM_*^2 ab^2 \, 3R^2}{32\zeta^3 [R^2 + (a+\zeta)^2]^{5/2}}, \quad (A9)$$

and

$$\Sigma_* \mathcal{S}_{\text{is}}^2 = \frac{GM_*^2 b^2 \, 3(a+\zeta)^2}{64\zeta^2 [R^2 + (a+\zeta)^2]^{5/2}}. \quad (A10)$$

In the case of the maximum rotation decomposition, applying (15) one obtains:

$$V_p^2(R,z) = \mathcal{V}_{\text{is}}^2(R,z) + \mathcal{S}_c^2(R,z), \quad (A11)$$

and

$$\sigma_p^2(R,z) = \mathcal{S}_s^2(R,z), \quad (A12)$$

with $\mathcal{S}_c^2 + \mathcal{S}_s^2 = \mathcal{S}_{\text{is}}^2$. After integration,

$$\Sigma_* \mathcal{S}_c^2 = \frac{GM_*^2 b^2}{64\zeta^2 (a+\zeta)^4} \times$$

$$R\left\{8 - \frac{8R^5 + 20(a+\zeta)^2 R^3 + 15(a+\zeta)^4 R}{[R^2+(a+\zeta)^2]^{5/2}}\right\}, (A13)$$

and

$$\Sigma_* \mathcal{S}_s^2 = \frac{GM_*^2 b^2}{64\zeta^2 (a+\zeta)^4} \times$$

$$\left\{\frac{8R^4 + 12(a+\zeta)^2 R^2 + 3(a+\zeta)^4}{[R^2+(a+\zeta)^2]^{3/2}} - 8R\right\}. \quad (A14)$$

Note that, because of the intrinsic underdeterminacy of the Jeans equations, $v_p^2$ has to be independent of the particular adopted velocity decomposition, and a check shows that in both of our cases $v_{\rm p}^2 = \mathcal{V}_{\rm is}^2 + \mathcal{S}_{\rm is}^2$.

Finally we compute the face-on projection of the velocity dispersion: after integration along the $z$ axis one obtains $\Sigma_*(0)\sigma_{\rm p}^2(0) = (GM_*^2/2\pi b^3)\widetilde{\Sigma}_*(s)\widetilde{\sigma}_{\rm p}^2(s)$, where

$$\widetilde{\Sigma}_*(s)\widetilde{\sigma}_{\rm p}^2(s) = \frac{\pi}{4s^4} +$$

$$\frac{3(8s^6 - 8s^4 + 7s^2 - 2)F(s) - s(26s^4 - 17s^2 + 6)}{12s^4(1-s^2)^3}, \quad (A15)$$

with $\widetilde{\Sigma}_*(0)\widetilde{\sigma}_{\rm p}^2(0) = 3\pi/32$, and $\widetilde{\Sigma}_*(1)\widetilde{\sigma}_{\rm p}^2(1) = \pi/4 - 16/21$.

## APPENDIX B: VIRIAL THEOREM FOR MN+MN MODELS

We compute here the analytical expressions for the terms entering the virial theorem of a MN density distribution that is interacting with another MN distribution. In order to simplify the notation and the resulting formulae, all the energies appearing in the virial theorem are normalized to the reference energy $\mathcal{E} = GM_*^2/8b$. From the linearity of the Jeans equations with respect to the potential, each of the contributions to the virial theorem is split in a part involving the interaction of one MN distribution with itself, and another involving the interaction with the other MN component; so $|W| = |W_*| + |W_{\rm h}|$, $T = T_* + T_{\rm h}$ and $\Pi = \Pi_* + \Pi_{\rm h}$. We start studying $|W|$, that is obviously independent of the assumed velocity decomposition:

$$|W_*| = -\frac{1}{2}\int \rho_* \Phi_* dV = \mathcal{E}\widetilde{W}_*(s), \quad (B1)$$

and after integration over the whole space, one gets

$$\widetilde{W}_*(s) = \frac{1-2s^2}{s(1-s^2)} - \frac{\pi}{2s^2} + \frac{F(s)}{s^2(1-s^2)} \quad (B2)$$

where $F(s)$ is given in (A1), and so $\widetilde{W}_*(0) = 3\pi/4$, and $\widetilde{W}_*(1) = 8/3 - \pi/2$. The interaction energy can be written as

$$|W_{\rm h}| = \int \rho_* \left[z\frac{\partial \Phi_{\rm h}}{\partial z} + R\frac{\partial \Phi_{\rm h}}{\partial R}\right] dV = \mathcal{E}\widetilde{W}_{\rm h}(c,s), \quad (B3)$$

with

$$\frac{(1-c^2)^2}{\mathcal{R}}\widetilde{W}_{\rm h}(c,s) =$$

$$\frac{\pi}{2}\left(c - \frac{1}{c}\right)^2 - \left(2c + \frac{1}{c}\right) + \left(4 - \frac{1}{c^2}\right)F(c) +$$

$$s\left[\left(2c^2 - 1 + \frac{2}{c^2}\right) - \pi\frac{(1-c^2)^2}{c^3} + \left(\frac{2}{c^3} - \frac{5}{c}\right)F(c)\right]; \quad (B4)$$

so $\widetilde{W}_{\rm h}(0,0)/\mathcal{R} = 3\pi/4$, and $\widetilde{W}_{\rm h}(1, 0 \leq s \leq 2)/\mathcal{R} = \pi/2 - 16/15 + s(56/15 - \pi)$.

The gravitational energy of the stellar distribution inside the dark matter halo potential is given by:

$$|U_{\rm h}| = -\int \rho_* \Phi_{\rm h} dV = \mathcal{E}\widetilde{U}_{\rm h}(c), \quad (B5)$$

which after integration becomes

$$\widetilde{U}_{\rm h}(c) = 2\mathcal{R}\widetilde{W}_*(c) \quad (B6)$$

where the function on the r.h.s. is the same given in (B2).

Moving now to the kinetic virial terms, and defining

$$T = \frac{1}{2}\int \rho_* \overline{v_\varphi}^2 dV, \quad (B7)$$

and

$$\Pi_{\rm is} = \int \rho_* \sigma^2 dV, \quad (B8)$$

after a simple algebraic manipulation of (13) and (14), we obtain for the Satoh's decomposition:

$$T = \mathcal{E}k^2\widetilde{T}, \quad (B9)$$

and

$$\Pi = \mathcal{E}[3\widetilde{\Pi}_{\rm is} + 2(1-k^2)\widetilde{T}]. \quad (B10)$$

For the maximum rotation,

$$T = \mathcal{E}(\widetilde{T} + 0.5\widetilde{\Pi}_{\rm is}), \quad (B11)$$

and

$$\Pi = \mathcal{E}2\widetilde{\Pi}_{\rm is}. \quad (B12)$$

So the analytic expressions for the two functions $\widetilde{T}$ and $\widetilde{\Pi}_{\rm is}$ are needed. Integrating by parts on $z$, we have

$$\Pi_{\rm is} = 4\pi \int_0^\infty z dz \int_0^\infty R\rho_* \frac{\partial \Phi_{\rm tot}}{\partial z} dR = \Pi_{\rm is,*} + \Pi_{\rm is,h}, \quad (B13)$$

and for the ordered kinetic energy, integrating by parts on $R$ and then on $z$:

$$T = 2\pi \int_0^\infty dz \int_0^\infty R^2 \rho_* \frac{\partial \Phi_{\rm tot}}{\partial R} dR - \Pi_{\rm is} \equiv T_0 - \Pi_{\rm is}. \quad (B14)$$

After tedious calculations, one obtains:

$$\widetilde{\Pi}_{\rm is,*}(s) = \frac{\pi}{2s^2} - \frac{1}{s(1-s^2)} + \frac{(2s^2-1)F(s)}{s^2(1-s^2)} \quad (B15)$$

with $\widetilde{\Pi}_{\rm is,*}(0) = \pi/4$, and $\widetilde{\Pi}_{\rm is,*}(1) = \pi/2 - 4/3$. A formally identical integration gives:

$$\widetilde{\Pi}_{\rm is,h}(c) = \mathcal{R}\widetilde{\Pi}_{\rm is,*}(c). \quad (B16)$$

Finally, for the ordered component $T_0$, we have

$$\widetilde{T}_{0,*}(s) = \frac{1}{s} - \frac{\pi}{2s^2} + \frac{F(s)}{s^2}, \quad (B17)$$

with $\widetilde{T}_{0,*}(0) = \pi/4$, $\widetilde{T}_{0,*}(1) = 2 - \pi/2$, and

$$\frac{(1-c^2)^2}{\mathcal{R}}\widetilde{T}_{0,\rm h}(c,s) = -\frac{3c}{2} + \left(c^2 + \frac{1}{2}\right)F(c) +$$

$$\frac{s}{2}\left[\left(2c^2 - 1 + \frac{2}{c^2}\right) - \frac{\pi(1-c^2)^2}{c^3} + \left(\frac{2}{c^3} - \frac{5}{c}\right)F(c)\right], (B18)$$

with $\widetilde{T}_{0,\rm h}(0,0)/\mathcal{R} = \pi/4$, and $\widetilde{T}_{0,\rm h}(1, 0 \leq s \leq 2)/\mathcal{R} = 2/15 + s(28/15 - \pi/2)$.

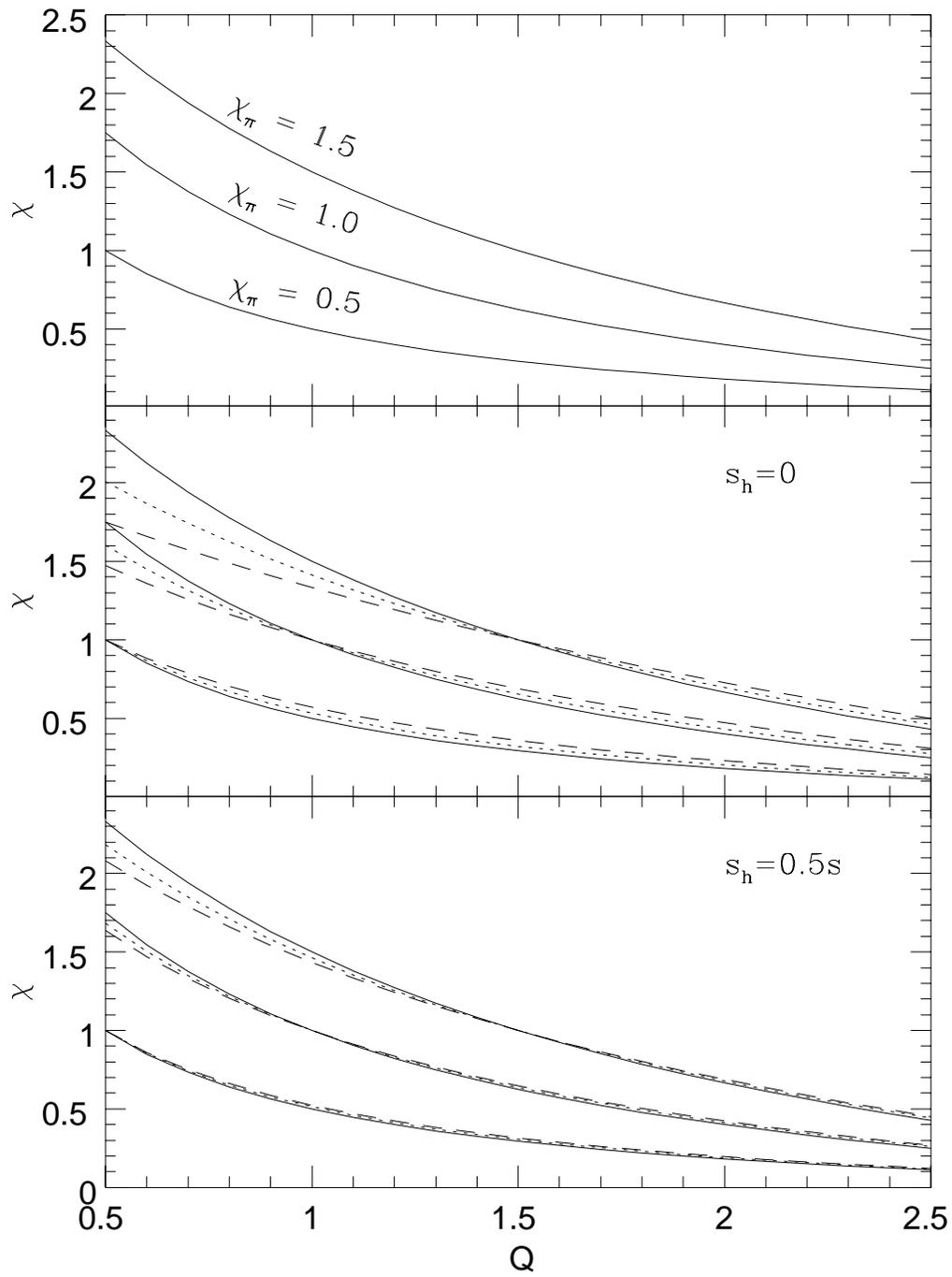

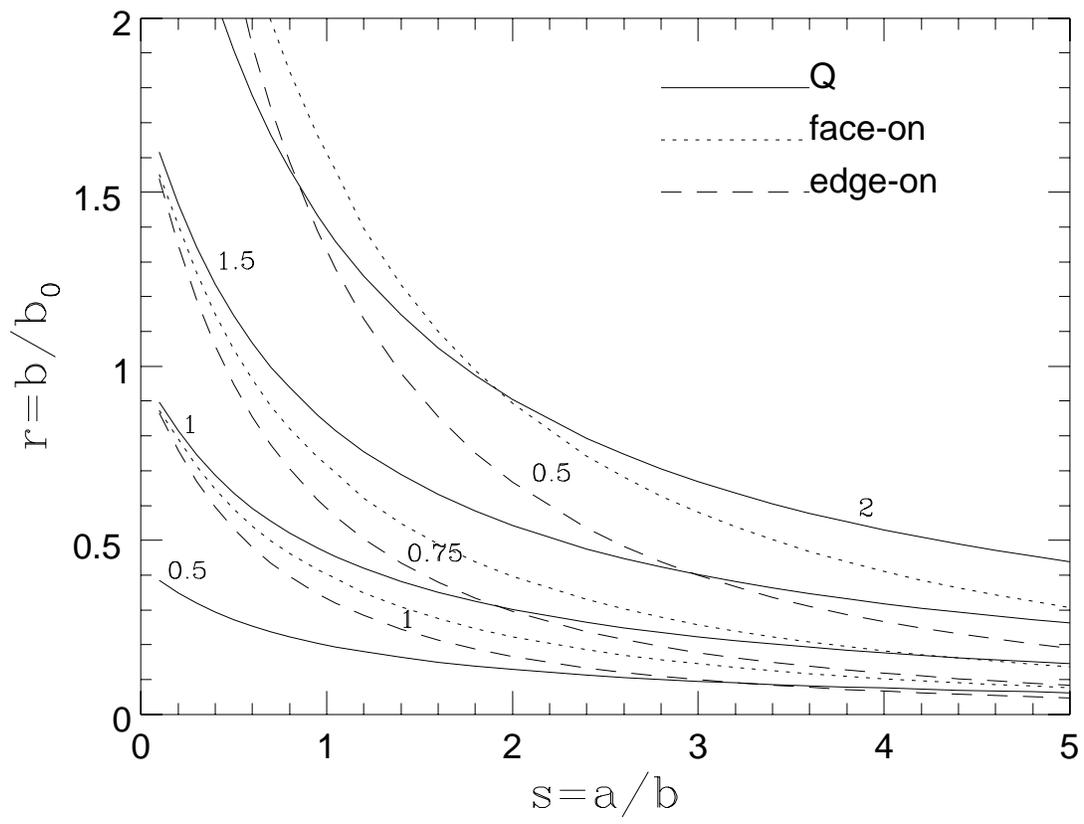

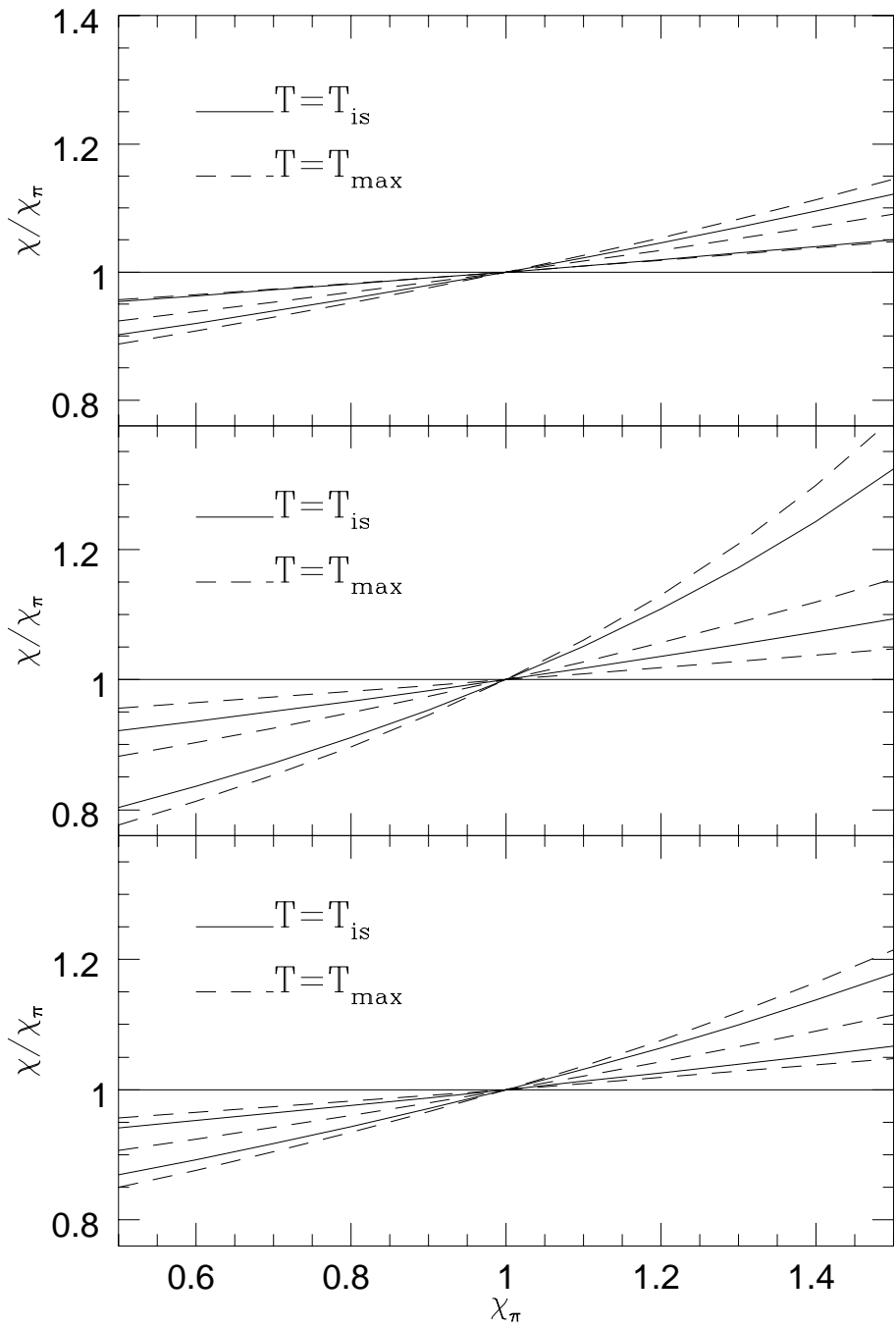

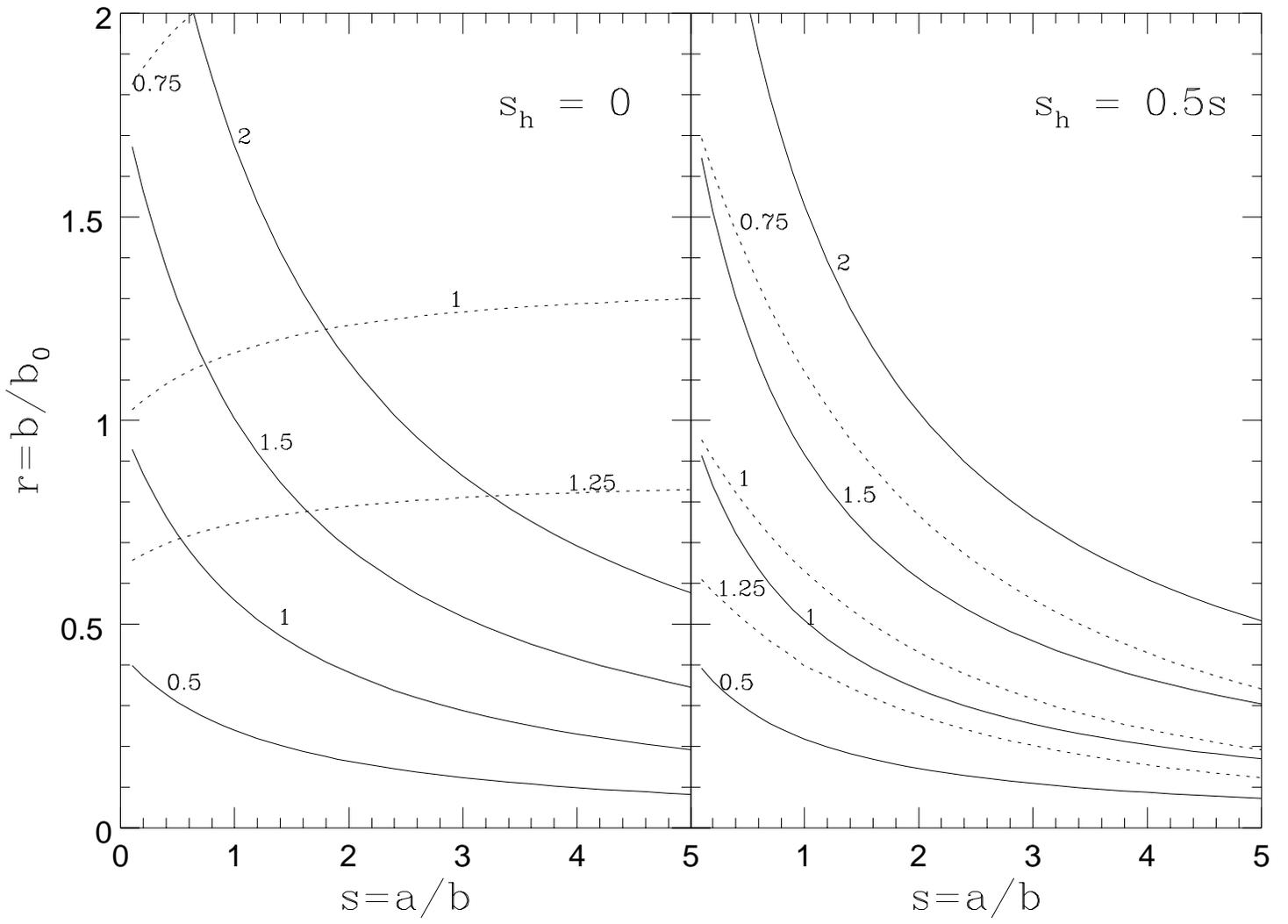